\documentclass[%
 reprint,
 amsmath,amssymb,
 aps,
]{revtex4-1}

\usepackage{graphicx}% Include figure files
\usepackage{dcolumn}% Align table columns on decimal point
\usepackage{bm}% bold math
\usepackage{amsmath}
\usepackage{enumerate}
\usepackage{setspace}
\usepackage{dsfont}
\usepackage[dvipdfm,  %pdflatex,   pdftex   这里决定运行文件的方式不同
            pdfstartview=FitH,
            CJKbookmarks=true,
            bookmarksnumbered=true,
            bookmarksopen=true,
            colorlinks, %注释掉此项则交叉引用为彩色边框(将colorlinks和pdfborder同时注释掉)
            pdfborder=001,   %注释掉此项则交叉引用为彩色边框
            linkcolor=blue,
            anchorcolor=blue,
            citecolor=blue
            ]{hyperref}
\begin{document}

\preprint{APS/123-QED}

\title{Composable security analysis of continuous-variable measurement-device-independent quantum key distribution with squeezed states for coherent attacks}% Force line breaks with \\

\author{Ziyang Chen,$^{1}$ Yichen Zhang,$^{2}$ Gan Wang,$^{1}$ Zhengyu Li,$^{1}$ and Hong Guo$^{1，＊}$}
\email{hongguo@pku.edu.cn}
\address{$^{1}$State Key Laboratory of Advanced Optical Communication System and Network, School of Electronics Engineering and Computer Science and Center for Quantum Information Technology, Peking University, Beijing 100871, China\\
$^{2}$State Key Laboratory of Information Photonics and Optical Communications, Beijing University of Posts and Telecommunications, Beijing 100876, China}

%\email{hongguo@pku.edu.cn}


\begin{abstract}
Measurement-device-independent quantum key distribution protocol, whose security analysis does not rely on any assumption on the detection system, can immune the attacking against detectors. We give a first composable security analysis for continuous-variable measurement-device-independent quantum key distribution using squeezed states against general coherent attacks. The security analysis is derived based on the entanglement-based scheme considering finite size effect. A version of entropic uncertainty relation is exploited to give a lower bound on the conditional smooth min-entropy by trusting Alice's and Bob's devices. The simulation results indicate that, in the universal composable security framework, the protocol can tolerate 2.5 dB and 6.5 dB channel loss against coherent attacks with direct and reverse reconciliation, respectively.

%Our work further improves the security analysis of CV-MDI QKD using squeezed states.
\end{abstract}

\pacs{Valid PACS appear here}% PACS, the Physics and Astronomy
\maketitle

%\tableofcontents

\section{Introduction}
Quantum key distribution (QKD) \cite{RevModPhys.74.145,RevModPhys.81.1301}, which is an indispensable part of today's quantum cryptography, allows two legitimate users (Alice and Bob) to distribute keys secretly thanks to quantum physics. The most attractive property of QKD may be the information-theoretic security against any potential attacks. Broadly speaking, QKD has two main approaches: one is discrete-variable (DV) QKD, and the alternative is continuous-variable (CV) QKD \cite{RevModPhys.77.513,RevModPhys.84.621,Entropy.17.6072}. Compared to DV-QKD protocols, CV-QKD protocols are based on variants of homodyne detection which is ``off-the-shelf'' \cite{Phys.Rev.Lett.88.057902.2002, Nature.421.238.2003, Phys.Rev.Lett.93.170504.2004}, and can perform high secret key rates for metropolitan range. Various novel CV-QKD protocols were proposed recent years, including two-way quantum cryptographic protocol \cite{Nat.Phys.4.726.2008, Int.J.Quantum.Inf.10.1250059.2012, J.Phys.B.47.035501.2014, Phys.Rev.A.92.062323.2015, Sci.Rep.6.22225.2016, J.Phys.B.At.Mol.Opt.Phys.50.035501.2017}, single-quadrature protocols \cite{Phys.Rev.A.92.062337.2015, Quantum.Inf.and.Comp.16.1081.2016}, floodlight QKD protocol \cite{Phys.Rev.A.94.012322.2016, Phys.Rev.A.95.012332.2017, arXiv.1712.04973.2017} and so forth, which enrich the field of CV-QKD. A new CV protocol design framework has been proposed to design protocols according to user's needs \cite{2018LI}, which can be achieved by arbitrary non-orthogonal states. Experiments \cite{Nat.Photonics.7.378, Lance_PhysRevLett_2005, Lodewyck_PhysRevA_2007, Qi_PhysRevA_2007, Khan_PhysRevA_2013, Nat.Photonics.8.333.2014}, especially field tests \cite{field_test} for distributing secret keys over long distances are currently achievable, making CV protocols competitive with respect to their DV counterparts.

The security-proof toolbox of CV-QKD has been enriched over the past few years, such as de Finetti theorem \cite{PhysRevLett.114.070501,Phys.Rev.Lett.118.200501}, postselection technique \cite{Phys.Rev.Lett.102.020504,PhysRevLett.110.030502}, the entropic uncertainty relations \cite{PhysRevLett109.100502,PhysRevA.90.042325,NatCommun6.8795} and so forth. Many protocols show their security against collective attacks via a Gaussian optimality arguement \cite{PhysRevLett.96.080502,PhysRevLett.97.190503,PhysRevLett.97.190502,PhysRevLett.101.200504} but are only considered in the asymptotic limit. Fortunately, those security-proof tools make it possible to generalize the security analysis to the most general coherent attacks even considering finite-size effect. For instance, under Gaussian modulation, the coherent state protocol with heterodyne detection was proved secure against coherent attacks with the help of rotation invariance \cite{PhysRevLett.114.070501}, and the squeezed state \cite{Phys.Rep.684.1.2017} protocol with homodyne detection is secure using entropic uncertainty relations \cite{PhysRevLett109.100502}.

Apart from theoretical security analysis, practical security analysis in QKD is gradually paid attentions to take the gap between theory and practice into consideration. Measurement-device-independent (MDI) QKD protocol is a genius idea to immune the attacking against detectors \cite{PhysRevLett.108.130502,PhysRevLett.108.130503,PhysRevA.89.052301,PhysRevA.90.052325,NaturePhotonics9.397,Entropy.17.4547} , moving towards practical security of QKD. CV-MDI QKD, as one of the candidate protocol to achieve multipartite communication \cite{PRA.93.022325.2016, arXiv.1709.06988.2017, PRA.97.032311.2018, arXiv.1605.05445.2016}, has been shown to against collective attacks and some work also take finite-size effect into account \cite{PhysRevA.96.042332,PhysRevA.96.042334, FiO.2017.JW4A.33}. Recently, the composable security analysis of CV-MDI QKD, which could be applied both to coherent-state protocols and to entangled-state protocols, has been proposed to defend general coherent attacks via Gaussian de Finetti reduction \cite{arXiv.1704.07924}, while the composable security analysis of that using squeezed states under coherent attacks has not been discussed yet.

It should note that the entropic uncertainty relations are paid a lot of attentions in both DV-QKD and CV-QKD's security proofs \cite{Nat.Commun.3.634,RevModPhys.89.015002}. There is a large family of entropic uncertainty relation, among which the infinite dimensional state-independent entropic uncertainty relation with quantum memories was studied in depth \cite{MathPhys.306.165} and it was soon applied for the security proof of squeezed-state protocol with homodyne detection \cite{PhysRevLett109.100502,PhysRevA.90.042325,NatCommun6.8795}. The entropic uncertainty method can be exploited to prove the security of squeezed-state CV-MDI QKD protocol directly.

In this paper, we use the similar method as Ref. \cite{PhysRevLett109.100502} did to the squeezed-state CV-MDI QKD protocol by trusting Alice's and Bob's devices, and show the performance against general coherent attacks, which is based on an state-independent entropic uncertainty relation with quantum side information for smooth entropies.	Meanwhile, the analysis not only considers the finite-size effect, but also takes some necessary steps into account, such as channel parameter estimation and error correction, so that the final secret key length has to be reduced due to the fact that those estimation phases inevitably consume amount of keys. Moreover, we analyze both direct and reverse reconciliation scenarios. Focusing on the extremely asymmetric cases, where Bob is placed on the Charlie's side, the ideal case (modulation variance tending to infinity) and a practical feasible parameters case (modulation variance as small as 5.04, referring to 10 dB squeezing \cite{Eberle:13}, with imperfect reconciliation efficiency $\beta  = 96.9\%$ \cite{Quantum.Inf.Comput.17.1123}) are both discussed at different block lengths. More general cases are also discussed in the appendix.

The paper is organized as follows. In Sec. \ref{Composable security definition in QKD}, a short review on the definition of composable security in QKD is described. In Sec. \ref{Description of squeezed-state CV-MDI QKD protocol}, we provide a detailed description of the squeezed-state CV-MDI QKD protocol against general coherent attacks under the entanglement-based scheme. In Sec. \ref{Uncertainty relation and secret key rate}, we introduce a version of state-independent entropy uncertainty relation conditional on quantum side-information into the security analyse of the protocol and derive the secure key rate against coherent attacks. Then, in Sec. \ref{SIMULATION RESULTS AND DISCUSSION}, we give out the simulation results of the secret key rate in both direct and reverse reconciliation cases, especially under extremely asymmetric scenarios. Finally, a summary of the paper is given in Sec. \ref{Conclusion}.

\section{Framework of the security analysis}

In this section, a brief introduction about the definition of composable security in QKD is given, and the details can be found in Ref. \cite{IEEE2001,NJP.11.085006}. Then the CV-MDI QKD protocol using squeezed states against coherent attacks is described in detail, followed by the entropic uncertainty relation to obtain the secret key rate of the protocol.

\subsection{Composable security definition}
\label{Composable security definition in QKD}

Roughly speaking, a protocol can be called `security', which should satisfy three criteria called `robustness', `correctness' and `secrecy'. If the probability of producing empty set of secret key is not higher than ${\varepsilon _{rob}}$ when eavesdropper is inactive, a protocol is called ${\varepsilon _{rob}}$-robust.

A QKD protocol can be called `correct', if Alice and Bob can get the same keys for any initial quantum state ${\Psi _{ABE}}$ (no matter what strategy of the adversary may be used to the quantum state). The secret key is denoted by ${S_A}$ and ${S_B}$ after they finish the protocol, and a protocol is called ${\varepsilon _{cor}} $-correct if the probability of producing different sets of secret key between ${S_A}$ and ${S_B}$ is not higher than ${\varepsilon _{cor}}$, i.e., $\Pr \left[ {{S_A} \ne {S_B}} \right] \le {\varepsilon _{cor}}$.

A final key is $\Delta$-secret if it is $\Delta$-close to a uniformly distributed key that is unpredictable for the adversary. Here $\Delta$ quantifies the distance between a practical key and an ideal one, for a $\Delta$-secret protocol, which should satisfy

\begin{equation}\
\frac{1}{2}{\left\| {{\rho _{S_AS_BE'}} - {\omega _l} \otimes {\rho _{E'}}} \right\|_1} \le \Delta,
\end{equation}
where ${{\rho _{S_AS_BE'}}}$ is the practical state mixing Alice, Bob and the potential adversary Eve's strings ${S_A}$, ${S_B}$ and $E'$, and ${\omega _l}$ is the fully mixed state on classical strings of length $l$. ${\omega _l} \otimes {\rho _{E'}}$ shows the ideal classical-quantum state is separable. Hence if $\Delta=0$ for any Eve's attack strategies, a QKD protocol can be called secret. Moreover, a protocol is called ${\varepsilon _{\sec }}$-secret, if it is ${\varepsilon _{\sec }}$-indistinguishable from a ideal secret protocol. In particular, a protocol is ${\varepsilon _{\sec }}$-secret, if it outputs $\Delta$-secure keys with $\left( {1 - {p_{abort}}} \right)\Delta  \le {\varepsilon _{\sec }}$, where ${{p_{abort}}}$ is the probability that the protocol aborts. ${\varepsilon _{rob}}$, ${\varepsilon _{cor}}$ and ${\varepsilon _{sec}}$ are parameters to qualify robustness, correctness and secrecy respectively and they will affect the final rate of secret key.

A QKD protocol is called secure if it satisfies both correct and secret. It is called ${\varepsilon } $-secure if it is ${\varepsilon } $-indistinguishable from a secure protocol. In particular, a protocol is ${\varepsilon } $-secure, if it is ${\varepsilon _{cor}} $-correct and ${\varepsilon _{\sec }}$-secret with ${\varepsilon _{cor}} + {\varepsilon _{\sec }} \le \varepsilon $.

\subsection{CV-MDI QKD using squeezed states against coherent attacks}
\label{Description of squeezed-state CV-MDI QKD protocol}
In this subsection, we describe the squeezed-state CV MDI QKD protocol for which we prove composable security against coherent attacks based on the entropic uncertainty relation. Here we focus on the entanglement-based (EB) model of the protocol \cite{Quantum.Inf.Comput.3.535} instead of the the prepare \& measure (PM) version, for the former scheme is often used in the security analysis of QKD, while the later is easy to implement, and once the security of the EB scheme is proved, the security of the PM version is easily obtained because of the equivalence between two schemes \cite{PhysRevA.89.052301}. The EB scheme of the squeezed-state CV-MDI QKD protocol (Fig. \ref{EB_protocol}) is described as follows:

 \begin{figure}[t]
  \centering
  \includegraphics[width=7.5cm]{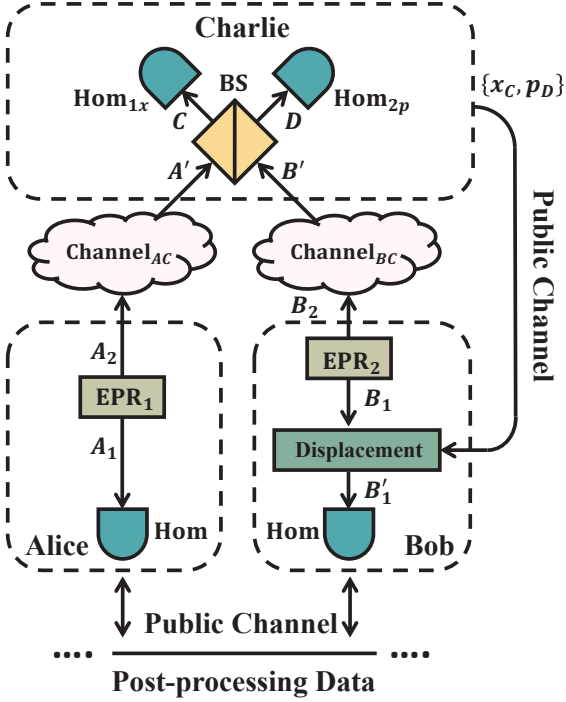}\\
  \caption{(Color online) EB scheme of the squeezed-state CV-MDI QKD protocol. EPR: two-mode squeezed state. Hom: homodyne detection. Hom$_{1x}$: homodyne detection of measuring the $x$-quadrature. Hom$_{2p}$: homodyne detection of measuring the $p$-quadrature. $X_C$ ($P_D$): measurement results of Hom$_{1x}$ (Hom$_{2p}$). BS: 50:50 balanced beam splitter. Channel$_{AC}$ (Channel$_{BC}$): totally untrusted quantum channel between Alice (Bob) and Charlie controlled by adversary. Public channel: authenticated channel using for classical communication.}
\label{EB_protocol}
\end{figure}
\begin{enumerate}[1.]
\item   \textbf{State Preparation:} Alice and Bob prepare $2N$ two-mode squeezed vacuum states EPR$_1$ and EPR$_2$ with variances ${V_A}$ and ${V_B}$, respectively. They keep mode ${A_1}$ and ${B_1}$ on each side and then send the other modes ${A_2}$ and ${B_2}$ to the untrusted third party (Charlie) through two insecure quantum channels, i.e. channel$_{AC}$ and channel$_{BC}$;

\item 	\textbf{Bell Measurement:} Charlie applies Bell detection of the received quantum states. Modes $A'$ and $B'$ are combined with a balanced beam splitter with output $C$ and $D$, and afterwards, $x$-quadrature of mode $C$ and $p$-quadrature of mode $D$ are measured with homodyne detectors. The results of joint measurement ${x_C}$ and ${p_D}$ are announced to Alice and Bob through public classical channel;

\item 	\textbf{Displacement:} After receiving Charlie's measurement results $\left\{ {{x_C},{p_D}} \right\}$, Bob apply local displacement operations $D\left( \beta  \right)$ on mode ${B_1}$ to get mode ${B_1}^\prime$, where $\beta  = g\left( {{x_C} + {p_D}} \right)$ and $g$ is the gain of this operation related to the total channel loss;

\item 	\textbf{Measurement:} Alice and Bob measure the $2N$ modes using homodyne detection which randomly detect the $x$-quadrature or $p$-quadrature and the measurement outcomes are discretized with the finite range analogue-to-digital converter (ADC). For every two modes ${A_1}$ and ${B_1}^\prime$, Alice gets the data $\left\{ {{X_A}\left( {{P_A}} \right)} \right\}$ and Bob gets the data $\left\{ {{X_B}\left( {{P_B}} \right)} \right\}$ respectively;

\item 	\textbf{Sifting:} Both of two communication parties announce which quadrature they choose through an authentic pubic channel. They hold the data which the selected quadratures are the same and discard the rest. The length of effective data after this step reduces to about $N$ in each party.

\item	\textbf{Channel Parameter Estimation:} Once Alice and Bob have collected sufficient correlated data, they use the public channel to perform parameter estimation to check the correlation between their data. The two parties randomly choose a common subset of length $k_{pe}$ from the sifted data and estimate the average distance between their samples:

\begin{equation}\
d\left( {X_A^{pe},X_B^{pe}} \right) = \frac{1}{k_{pe}}\sum\nolimits_{i = 1}^{k_{pe}} {\left| {X_{A,i}^{pe} - X_{B,i}^{pe}} \right|},
\end{equation}
where $X_A^{pe} = \left( {X_{A,i}^{pe}} \right)_{i = 1}^{k_{pe}}$ and $X_B^{pe} = \left( {X_{B,i}^{pe}} \right)_{i = 1}^{k_{pe}}$. If $d\left( {X_A^{pe},X_B^{pe}} \right)$ is smaller than a certain parameter ${d_0}$, they proceed and otherwise, the protocol aborts. The parameter ${d_0}$ is the distance between the measurement results of Alice and Bob, which should be chosen small enough to ensure the data are correlated enough. Data $X_A^{pe}$ and $X_B^{pe}$ are also used to estimate the amount of information needed in Error correction step;

\item	\textbf{Error Correction:} Alice sends some information to Bob and Bob corrects the errors in his data using error reconciliation algorithm (direct reconciliation), or Alice corrects the errors in her data with the help of Bob's sending information (reverse reconciliation). It may cost length of ${\ell _{EC}}$ secret keys during error correction phase. After that, two parties do the hash check \cite{PhysRevLett.114.070501}, i.e. they expend the length of $k_{check}$ extra data to check if both hashes coincide. If this check passes, the protocol resumes, otherwise it aborts;

\item	\textbf{Calculation of Secret Key Length:} Alice and Bob calculate the secret key length $\ell$ according to the presented secret key length formula and entropic uncertainty relation which will be shown in Sec. \ref{Uncertainty relation and secret key rate}. If the secret key length is negative, they abort the protocol;

\item	\textbf{Privacy Amplification:} Both of two communication parties apply a hash function \cite{J.Comput.Syst.Sci.18.143} on their corrected strings respectively to generate the secret key of length $\ell $.
\end{enumerate}

\subsection{Uncertainty relation and secret key rate}
\label{Uncertainty relation and secret key rate}

In previous researches of CV-QKD, in general, a practical homodyne detector is modeled as an ideal homodyne detector and an ADC with finite range \cite{PhysRevLett.114.070501}. To illustrate the measurement phase in our protocol more clearly, without loss of generality, we model the homodyne detector as an ideal homodyne detector followed by an ADC with finite range and divide the measurement process into two steps. First, Alice and Bob use ideal homodyne detectors to measure the quadratures of the states that they received (${\rho _{{A_1}}}$ and ${\rho _{{{B}_1}'}}$ in Fig. \ref{EB_protocol}). The outputs of ideal homodyne detectors $(\left\{ {{Q_A},{P_A}}\right\}$, $\left\{ {{Q_B},{P_B}} \right\})$ in two sides are ideal continuous variables with infinite range, and the statistical distribution of each outcome should generally follow a Gaussian distribution.

It is important for the protocol to have high correlations between two parties' outcomes. However, due to the channel losses, the quadratures $x$ and $p$ at Alice and Bob sides $(\left\{ {{Q_A},{P_A}}\right\}$,$\left\{ {{Q_B},{P_B}} \right\})$ will decay. In order to handle that, the quadrature measurements in one of two parties $\left\{ {{Q_A},{P_A}}\right\}$ or $\left\{ {{Q_B},{P_B}} \right\}$ needs to be rescaled before grouping into the intervals. We use the transformations below (using Alice as an example):

\begin{equation}\
{Q_A} \to {\tilde Q_A} = {t_q}{Q_A},{P_A} \to {\tilde P_A} = t_p{P_A},
\label{rescale}
\end{equation}
where $t_q$ and $t_p$ denote the rescaling factors related to the channel loss of Channel$_{AC}$ and Channel$_{BC}$ (see Appendix \ref{Estimation of ${t_q}$ and ${t_p}$ in measurement stage} about the estimation). After that, the data between Alice and Bob should be correlated enough.

In step 2, Alice and Bob use ADCs with finite sampling range and finite resolution, to discretize the continuous quadratures $\left\{ {{{\tilde Q}_A},{{\tilde P}_A}} \right\}$ and $\left\{ {{Q_B},{P_B}} \right\}$ into different intervals: $\left( { - \infty , - \alpha } \right]$, $\left( { - \alpha   , - \alpha  + \delta } \right]$, $....$, $\left( {\alpha   ,\infty } \right)$. Here, $\alpha $ is the maximum discretization range of ADC, which takes the finite range of detectors into consideration in the security proof, and $\delta$ denotes the precision of the measurement. The corresponding outcome alphabet is denoted by $\chi  = \left\{ {1,2,...,{{2\alpha } \mathord{\left/{\vphantom {{2\alpha } \delta }} \right.\kern-\nulldelimiterspace} \delta }} \right\}$, where we assume ${{2\alpha } \mathord{\left/{\vphantom {{2\alpha } \delta }} \right.\kern-\nulldelimiterspace} \delta } \in \mathbb{N}$ and every measurement outcome corresponds to one of the intervals. Therefore, the data $\left\{ {{X_A}, {P_A}} \right\}$ in Alice side is obtained by discreting the quadrature measurements $\left\{ {{{\tilde Q}_A},{{\tilde P}_A}} \right\}$, likewise the data in Bob.

It should note that practical homodyne detection may lead to security problems since its outputs lack information of the quadratures. For instance, in equal-length intervals $\left( { - \alpha   , - \alpha  + \delta } \right]$, $....$, $\left( {  \alpha  - \delta ,  \alpha  } \right]$, owning to the finite sampling bits, any measurement outcomes inside of one sampling interval will map to the same value and it may cause lacking of the details about the state within each sampling interval, for one cannot determine whether the distribution of measured states is Gaussian distribution or other non-Gaussian distribution. Moreover, we assume that any information in another two infinite-length intervals $\left( { - \infty , - \alpha } \right]$ and $\left[ {\alpha ,\infty } \right)$ will also map to one value as a result of the finite sampling range property of ADC, e.g. one cannot distinguish whether the measured pulse is low energy pulse or high energy pulse, which makes the measurement outcomes short of the information about the state outside the range. Those imperfect feature of detection may open the loophole to potential Eve and a number of attacks, such as large energy attack, may be exploited to reduce the security of the protocol. There are in general two approaches to handle that problem. One is using the method as Ref. \cite{PhysRevLett109.100502} did by trusting Alice's and Bob's devices, and another solution is adding the energy test to provide detailed information about measured states (as Ref. \cite{PhysRevA.90.042325} did) to replace the trusted source assumption. This paper follows the former solution and assume that Alice and Bob produce trusted states with quadratures being larger than $\alpha$ with very small probability $p_{\alpha}$.

After the measurement step is done, the physical steps of the protocol are finished, and the rest of the protocol is treated as the post-processing part aiming at extracting secure keys from the raw keys. Due to the leftover hash lemma, the ${\varepsilon _c}$-correct and ${\varepsilon _s}$-secret key of length $\ell$ can be extracted \cite{arXiv.0512258.2006}, which satisfies
\begin{equation}\
\ell  \le H_{\min }^\varepsilon {\left( {{X_A}|E} \right)_\omega } - {\ell _{EC}} - O\left( {\log \frac{1}{{{\varepsilon _s}{\varepsilon _c}}}} \right),
\end{equation}
where ${\ell _{EC}}$ denotes the leakage information in error correction phase, and $H_{\min }^\varepsilon \left(  \cdot  \right)$ is the smooth min-entropy with smoothing parameter $\varepsilon$. $H_{\min }^\varepsilon \left( {{X_A}|E} \right)$ is the smooth conditional min-entropy of data ${X_A}$ conditioned on the information Eve may have, which quantifies Eve's uncertainty about the Alice's measurement outcomes. In the coherent attacks cases, the goal is to bound the smooth min-entropy $H_{\min }^\varepsilon \left( {{X_A}|E} \right)$ conditioned on the event that the protocol does not abort. Different from the parameter estimation method in Ref. \cite{arXiv.1704.07924}, the smooth min-entropy $H_{\min }^\varepsilon \left( {{X_A}|E} \right)$ can be estimated with the help of the entropic uncertainty relation conditioned on side information with infinite dimensional quantum memories \cite{PhysRevLett109.100502} in our paper.

Entropic uncertainty relations are used in some security proofs of QKD protocols giving its
power to describe the bounds of guessing uncertainty Eve may have, when both Alice and Bob perform measurements in two random bases in a certain tripartite quantum system. There is a large family of entropy uncertainty relations with both infinite-dimensional and finite-spacing formulas \cite{RevModPhys.89.015002}. However, a more operational way to express uncertainty is in terms of the discrete Shannon entropy rather than differential relations, so we follow above to calculate the secret key length with discrete Shannon entropy's version of uncertainty relation, and quantum side information is considered with smooth min- and max- entropies.

The scenario of uncertainty relations can be understood as follows: The tripartite state ${\omega _{ABE}}$ with Alice, Bob and Eve hold infinite dimensional quantum systems $A$, $B$ and $E$ respectively. Alice randomly measures quadrature $x$ or $p$ on state ${\omega _A} = Tr{_{BE}}\left[ {{\omega _{ABE}}} \right]$ in each run and stores the outcomes in one of two classical systems. The same operation is done at Bob side acting at ${\omega _B} = T{r_{AE}}\left[ {{\omega _{ABE}}} \right]$. The outcome strings are denoted by $\left\{ {{X_A},{P_A}} \right\}$ and $\left\{ {{X_B},{P_B}} \right\}$ respectively. After sifting, two pairs of random strings $\left\{ {{X_A},{X_B}} \right\}$ and $\left\{ {{P_A},{P_B}} \right\}$ should obey the uncertainty principle and Eve cannot predict Alice and Bob's measurement outcomes precisely. Hence the relation between smooth min- and max- entropies satisfies

\begin{equation}\
H_{\min }^\varepsilon {\left( {{X_A}|E} \right)_\omega } \ge n\log \frac{1}{{c\left( \delta  \right)}} - H_{\max }^{\varepsilon '}{\left( {{X_A}|{X_B}} \right)_\omega }.
\label{inequality_uncertainty}
\end{equation}

Here we assume the random selection is identically and independently distributed. The term $c\left( \delta  \right)$ is the `incompatibility' of the measurement operators and $H_{\max }^{\varepsilon '}{\left( {{X_A}|{X_B}} \right)_\omega }$ is the smooth max-entropy between the data of Alice and Bob with smoothing parameter $\varepsilon '$, which reads

\begin{equation}\
\varepsilon ' = {{{\varepsilon _s}} \mathord{\left/
 {\vphantom {{{\varepsilon _s}} {\left( {4{p_{pass}}} \right) - {{2f\left( {{p_\alpha },n} \right)} \mathord{\left/
 {\vphantom {{2f\left( {{p_\alpha },n} \right)} {\sqrt {{p_{pass}}} }}} \right.
 \kern-\nulldelimiterspace} {\sqrt {{p_{pass}}} }}}}} \right.
 \kern-\nulldelimiterspace} {\left( {4{p_{pass}}} \right) - {{2f\left( {{p_\alpha },n} \right)} \mathord{\left/
 {\vphantom {{2f\left( {{p_\alpha },n} \right)} {\sqrt {{p_{pass}}} }}} \right.
 \kern-\nulldelimiterspace} {\sqrt {{p_{pass}}} }}}}.
\end{equation}
with $f\left( {{p_\alpha },n} \right) = \sqrt {2\left( {1 - {{\left( {1 - {p_\alpha }} \right)}^n}} \right)}$ \cite{Supplemental_Material}, which is the function considering about the probability of the event in outside of the detection range $\left[ { - \alpha ,\alpha } \right]$. $c\left( \delta  \right)$ takes the measurement discretization into consideration, which is

\begin{equation}\
c\left( \delta  \right) = \frac{1}{{2\pi }}{\delta ^2} \cdot S_0^{\left( 1 \right)}{\left( {1,\frac{{{\delta ^2}}}{4}} \right)^2},
\end{equation}
where $S_0^{\left( 1 \right)}$ denotes the ${0^{th}}$ radial prolate spheroidal wave function of the first kind \cite{J.Math.Phys.51.072105}. $c\left( \delta  \right)$ can be well approximated with $c\left( \delta  \right) \approx {{{\delta ^2}} \mathord{\left/ {\vphantom {{{\delta ^2}} {\left( {2\pi } \right)}}} \right.\kern-\nulldelimiterspace} {\left( {2\pi } \right)}}$ when the length of interval $\delta $ is small. For a certain value of $\delta $, $c\left( \delta  \right)$ is a constant either, so the value of smooth min-entropy $H_{\min }^\varepsilon \left( {{X_A}|E} \right)$ can be estimated by upper bounding the smooth max-entropy $H_{\max }^{\varepsilon '}\left( {{X_A}|{X_B}} \right)$ between random strings ${X_A}$ and ${X_B}$.

To estimate the upper bound of $H_{\max }^{\varepsilon '}\left( {{X_A}|{X_B}} \right)$, the correlation of the data between Alice and Bob needs to be qualify first. Alice and Bob randomly choose a subset ${\chi ^{k_{pe}}}$ with string length $k_{pe}$ to calculate the average distance $d\left( {X_A^{pe},X_B^{pe}} \right)$ between their data $X_A^{pe}$ and $X_B^{pe}$ in parameter estimation step, where $pe$ stand for parameter estimation. If $d\left( {X_A^{pe},X_B^{pe}} \right) < {d_0}$, the $\varepsilon '$-smooth max-entropy can be bounded by

 \begin{equation}\
H_{\max }^{\varepsilon '}\left( {{X_A}|{X_B}} \right) \le n\log \gamma \left( {d\left( {{X_A},{X_B}} \right)} \right),
 \end{equation}
where $\gamma $ is a function arising from a large deviation consideration, which reads

  \begin{equation}\
 \gamma \left( t \right) = \left( {t + \sqrt {{t^2} + 1} } \right){\left[ {{t \mathord{\left/
 {\vphantom {t {\left( {\sqrt {{t^2} + 1}  - 1} \right)}}} \right.
 \kern-\nulldelimiterspace} {\left( {\sqrt {{t^2} + 1}  - 1} \right)}}} \right]^t}.
  \end{equation}

Using sampling theory, the quantity $d\left( {{X_A},{X_B}} \right)$ can be estimated by $d\left( {X_A^{pe},X_B^{pe}} \right)$ plus a correction $\mu $ with high probability. $\mu $ quantifies its deviation to $d\left( {{X_A},{X_B}} \right)$ considering about the finite-size statistical fluctuation. Finally, the $\ell $ length secret key can be extracted from the remaining data ${X_A},{X_B} \in {\chi ^n}$ with the length of $n$, which is written as

\begin{equation}\
\ell  = n\left[ {\log \frac{1}{{c\left( \delta  \right)}} - \log \gamma \left( {{d_0} + \mu } \right)} \right] - {\ell _{EC}} - \log \frac{1}{{\varepsilon _s^2{\varepsilon _c}}},
\label{keyrate}
\end{equation}
and
\begin{equation}\
\mu  = \frac{{2\alpha }}{\delta }\sqrt {\frac{{N\left( {{k_{pe}} + 1} \right)}}{{nk_{pe}^2}}\ln \frac{1}{{\varepsilon '}}}.
\label{data_fluctuation}
\end{equation}

Here the remaining data has a length $n = N - k_{pe} - k_{check}$ approximately, for some raw keys were cut off from the test steps above.

There are two main elements of Eq. (\ref{keyrate}), one is the estimation of smooth min-entropy $H_{\min }^\varepsilon \left( {{X_A}|E} \right)$, and the other is the leakage information ${\ell _{EC}}$ during error correction. The former, as mentioned above, can be estimated by the date $X_A^{pe}$ and $X_B^{pe}$, which is independent to the reconciliation methods \cite{note}, while the later is determined by the information reconciliation, hence Eq. (\ref{keyrate}) can be exploited to calculate secret key rate in both direct and reverse reconciliation cases.

For the direct reconciliation case, the leakage information in error correction step reads

\begin{equation}\
\ell _{EC}^{DR} = H\left( {{X_A}} \right) - \beta I\left( {{X_B}:{X_A}} \right),
\end{equation}
and in reverse reconciliation case it reads

\begin{equation}\
\ell _{EC}^{RR} = H\left( {{X_B}} \right) - \beta I\left( {{X_B}:{X_A}} \right),
\end{equation}
where $H\left( {{X_A}} \right)$ and $H\left( {{X_B}} \right)$ denote the discrete Shannon entropies, and $I\left( {{X_B}:{X_A}} \right)$ is the mutual information between Alice and Bob.

\section{Numerical simulation and discussion}
\label{SIMULATION RESULTS AND DISCUSSION}

In this section, we focus on the simulation results of the squeezed-state CV-MDI QKD protocol in the ideal detection case against coherent attacks. Section \ref{Uncertainty relation and secret key rate} illustrates that the simulation of secret key rate in our protocol can be divided into two parts, one is the estimation of smooth min-entropy $H_{\min }^\varepsilon \left( {{X_A}|E} \right)$ considering finite-size effect, the other is the leakage information ${\ell _{EC}}$ in error correction, which could be calculated with the help of the covariance matrix. Only the extremely asymmetric cases are discussed here as the examples, where Bob is located at Charlie's side ($T_{BC}=0$), for the transmission distance can reach the maximum \cite{PhysRevA.90.052325}. The discussion of symmetric cases can be seen in Appendix \ref{The protocol under different transmission losses} considering more general attack strategy.

Considering the EB version of the squeezed-state CV-MDI QKD protocol (Fig. \ref{EB_protocol}), the covariance matrix can be estimated by Alice and Bob's data directly in experiment, and here, without loss of generality, we assume that channels$_{AC}$ and channels$_{BC}$ are under two independent entangling cloner attacks to estimate the covariance matrix. We should point out that Eve's attack described here is not the optimal one \cite{NaturePhotonics9.397, Phys.Rev.A.91.022320.2015}. The entangling cloner attack is usually used to model a Gaussian channel affected by the environment and is analyzed to get a sense of a protocol's performance in experiment \cite{PhysRevA.90.052325}, and in experiment, we can calculate the amount of information used in error correction phase in parameter estimation step without assuming which attack Eve may use. Moreover, the estimation of leakage information does affect the final secret key rate, but not induce statistical fluctuation introduced by parameter estimation step, and all the statistical fluctuation introduced by parameter estimation has been considered in the estimation of max-entropy. Detailed derivation of the covariance matrix can be see in Appendix \ref{Covariance matrixes and calculating the leakage information in error correction}. First, Alice and Bob generate two-mode squeezed state ${\rho _{{A_1}{A_2}}}$ and ${\rho _{{B_1}{B_2}}}$ respectively. The covariance matrixes ${\gamma _{{A_1}{A_2}}}$ and ${\gamma _{{B_1}{B_2}}}$ read

\begin{equation}\
{\gamma _{{A_1}{A_2}}} = \left( {\begin{array}{*{20}{c}}
{{V_A}{\mathbb{I}_2}}&{\sqrt {V_A^2 - 1} {\sigma _z}}\\
{\sqrt {V_A^2 - 1} {\sigma _z}}&{{V_A}{\mathbb{I}_2}}
\end{array}} \right),
\end{equation}

\begin{equation}\
{\gamma _{{B_1}{B_2}}} = \left( {\begin{array}{*{20}{c}}
{{V_B}{\mathbb{I}_2}}&{\sqrt {V_B^2 - 1} {\sigma _z}}\\
{\sqrt {V_B^2 - 1} {\sigma _z}}&{{V_B}{\mathbb{I}_2}}
\end{array}} \right),
\end{equation}
where ${\mathbb{I}_2}$ is the identity matrix, ${V_{A\left( B \right)}}$ stand for the variance of Alice (Bob)'s two-mode squeezed state from Eve's view and ${\sigma _z} = \left( {\begin{array}{*{20}{c}} 1&0\\0&{ - 1}\end{array}} \right)$.

Before Charlie applies Bell measurement to mode $C$ and mode $D$, the whole state ${\rho _{{A_1}CD{B_1}}}$ can be described by a $8 \times 8$ covariance matrix ${\gamma _{{A_1}CD{B_1}}}$. Then mode $A'$ and $B'$ received by Charlie interfere at a beam splitter (BS) with two output $C$ and $D$ modes measured by homodyne detections respectively. The measurement results ${x_C}$ and ${p_D}$ are announced by Charlie in a public channel so that Bob can displace mode ${B_1}$ to ${B'_1}$. It is easy to get the covariance matrix ${\gamma _{{A_1}{B_1}^\prime }}$ of the state ${\rho _{{A_1}{B_1}^\prime }}$ shared by Alice and Bob, which reads

\begin{equation}\
{\gamma _{{A_1}{B_1}^\prime }} = \left( {\begin{array}{*{20}{c}}
{{V_A}{{\mathbb {I}}_2}}&{\sqrt {T\left( {V_A^2 - 1} \right)} {\sigma _z}}\\
{\sqrt {T\left( {V_A^2 - 1} \right)} {\sigma _z}}&{\left[ {T\left( {{V_A} - 1} \right) + 1 + Te} \right]{{\mathbb {I}}_2}}
\end{array}} \right),
\label{gammaAB}
\end{equation}
where

\begin{equation}\
T = \frac{{{T_1}}}{2}{g^2}.
\end{equation}

$T$ stands for the equivalent channel transmittance between Alice and Bob, ${T_1}$ is the channel transmittance between Alice and Charlie, and $g$ is the gain of displacement. The equivalent excess noise $e$ is given by

\begin{align}\
e &= 1 + \frac{1}{{{T_1}}}\left[ {2 + {T_2}\left( {{\varepsilon _2} - 2} \right) + {T_1}\left( {{\varepsilon _1} - 1} \right)} \right]          \notag\\
&+ \frac{1}{{{T_1}}}{\left( {\frac{{\sqrt 2 }}{g}\sqrt {{V_B}-1}  - \sqrt {{T_2}} \sqrt {{V_B} + 1} } \right)^2}.
\end{align}

In the numerical simulation, one can select $g = \sqrt {\frac{2}{{{T_2}}}} \sqrt {\frac{{{V_B}-1}}{{{V_B} + 1}}}$ so that the equivalent excess noise $e$ is optimal \cite{PhysRevA.89.052301}. Therefore, we can get

\begin{equation}\
e = {\varepsilon _1} + \frac{1}{{{T_1}}}\left[ {{T_2}\left( {{\varepsilon _2} - 2} \right) + 2} \right].
\end{equation}

Accordingly, the discrete Shannon entropies $H\left( {{X_A}} \right)$ and $H\left( {{X_B}} \right)$ have the following forms when $\delta$ is small (see Appendix \ref{Derivation of discrete Shannon entropy} about the detailed derivation):

\begin{equation}\
H\left( {{X_A}} \right) \approx \log \left( {\sqrt {2\pi e{V_A}} } \right) - \log \left( \delta  \right),
\end{equation}
and
\begin{equation}\
H\left( {{X_B}} \right) \approx \log \left( {\sqrt {2\pi e{V_B}^\prime } } \right) - \log \left( \delta  \right),
\end{equation}
where ${V_B}^\prime  = T\left( {{V_A} - 1} \right) + 1 + Te$. The mutual information between Alice and Bob can be well approximated, which reads

\begin{equation}\
I\left( {{X_A}:{X_B}} \right) \approx \frac{1}{2}\log \left( {\frac{{{V_A} + \chi }}{{\chi  + \frac{1}{{{V_A}}}}}} \right),
\end{equation}
where $\chi  = \frac{1}{T} - 1 + e$. Once we have obtained the form of covariance matrix in the EB model, with the help of Eq. (\ref{keyrate}), the secret key rate against coherent attack in both direct reconciliation and reverse reconciliation cases can be calculated.

\subsection{Direct reconciliation protocol}

\begin{figure}[t]
  \centering
  \includegraphics[width=7.5cm]{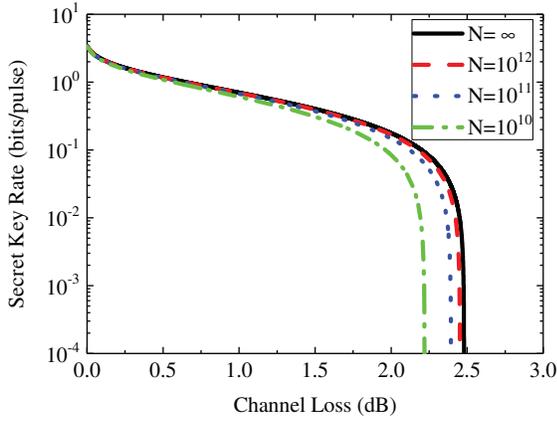}\\
  \caption{(Color online) Secret key rates of squeezed-state CV-MDI QKD protocol against coherent attack in the extremely asymmetric cases (${T_{BC}} = 0$) with direct reconciliation in the frame of composable security. Those lines are under the ideal conditions with ideal modulation variances ${V_A} = {V_B} = {10^5}$ and perfect reconciliation efficiency $\beta  = 1$. The block lengths from left to right curves show $N = {10^{10}}$ (green dot-dashed line), ${10^{11}}$ (blue dot line), ${10^{12}}$ (red dashed line) and $ \infty $ (black solid line), respectively. Here the discretization parameter is set to $d=13$, the excess noise ${\varepsilon _1} = {\varepsilon _2} = 0.002$, and the overall security parameter is smaller than ${10^{ - 20}}$.}
\label{ideal_asymmetric_D}
\end{figure}

\begin{figure}[t]
  \centering
  \includegraphics[width=7.5cm]{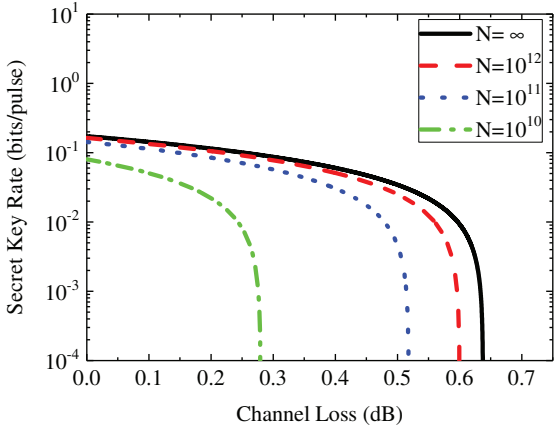}\\
  \caption{(Color online) Secret key rates of squeezed-state CV-MDI QKD protocol against coherent attack in the extremely asymmetric cases (${T_{BC}} = 0$) with direct reconciliation. The protocol with practical modulation variances ${V_A} = {V_B} = 5.04$ and imperfect reconciliation efficiency $\beta  = 96.9\% $ is considered. The block lengths from left to right curves are $N = {10^{10}}$ (green dot-dashed line), ${10^{11}}$ (blue dot line), ${10^{12}}$ (red dashed line) and $ \infty $ (black solid line), respectively. The discretization parameter, excess noises, and security parameters are chosen as in the case of ideal modulation.}
\label{practical_asymmetric_D}
\end{figure}

\begin{figure}[t]
  \centering
  \includegraphics[width=7.5cm]{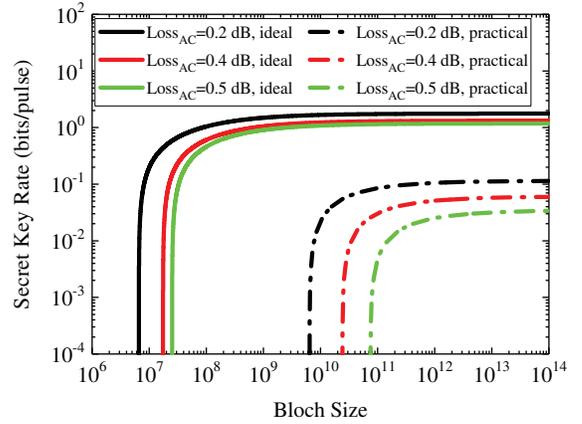}\\
  \caption{(Color online) Secret key rates vs block size for the extremely asymmetric case (${T_{BC}} = 0$) with direct reconciliation. The solid lines are under the ideal condition where modulation variances ${V_A} = {V_B} = {10^5}$ and perfect reconciliation efficiency $\beta  = 1$. The dot-dashed lines are under the practical condition where practical modulation variances ${V_A} = {V_B} = {5.04}$ and imperfect reconciliation efficiency $\beta  = 96.9\% $. From left to right, the transmittance of the quantum channel corresponds to loss of 0.2 dB (black line), 0.4 dB (red line), and 0.5 dB (green line), respectively.}
\label{N_change_D}
\end{figure}

First, numerical simulations of the secret-key rate in the direct reconciliation cases are performed. The performance of the extremely asymmetric structure with ideal modulation variances $\left( {{V_a} = {V_b} = {{10}^5}} \right)$ is given in Fig. \ref{ideal_asymmetric_D}. The perfect reconciliation efficiency ($\beta  = 1$) is set to get the optimal performance of this protocol against coherent attacks. The interval parameter is set to $\alpha  = 52$ \cite{PhysRevLett109.100502}, the discretization parameter $d=13$, the excess noises ${\varepsilon _1} = {\varepsilon _2} = 0.002$, and the overall security parameter is smaller than ${10^{ - 20}}$. The block length of information reconciliation $k_{pe}$ can be optimized in experiment. If $k_{pe}$ is too large, the final key rate may decrease due to a small quantity of raw key using for generating secret keys. On the contrary, one may not get accurate estimation of the channel parameters if $k_{pe}$ is too small. In this simulation, we choose the block length of information reconciliation about $1/10$ of the total length, i.e. $k_{pe}=N/10$. It can be seen that, when the block length is infinite-size ($N = \infty$), the protocol reach the longest transmission distance, with a corresponding channel loss of about 2.5 dB. In $N = 10^{12}$ case, the protocol is closed to the asymptotic rate.

The realistic performance is described under the condition that the practical variances are ${V_A} = {V_B} = 5.04$ (referring to 10 dB squeezing) and imperfect reconciliation efficiency is set to $\eta  = 96.9\% $. The key rate of a realistic extremely asymmetric case of the CV-MDI QKD protocol is described in Fig. \ref{practical_asymmetric_D}. We plot the key rate as a function of the channel loss $T_{AC}$, while the channel loss $T_{BC}$ is set to 0 dB, with different block lengths of $ {10^{10}}$, $ {10^{11}}$, $ {10^{12}}$ and infinite-size. For the asymptotic case $N \to \infty $, the maximum tolerable channel loss can reach approximately 0.64 dB (black solid line), which shows a distance between practical and ideal cases. The practical performance can be optimized using squeezed states with higher squeeze factor \cite{Opt.Let.43.110}.

What's more, for given distances, we plot the secret key rate vs. the block size when both ideal and practical parameters are given (Fig. \ref{N_change_D}). The channel losses are 0.2 dB, 0.4 dB, and 0.5 dB respectively. When the block length reduces, the secret key rate decreases rapidly and one can not generate secret key when the block length is smaller than ${10^{10}}$ under the practical parameters.

\subsection{Reverse reconciliation protocol}

\begin{figure}[t]
  \centering
  \includegraphics[width=7.5cm]{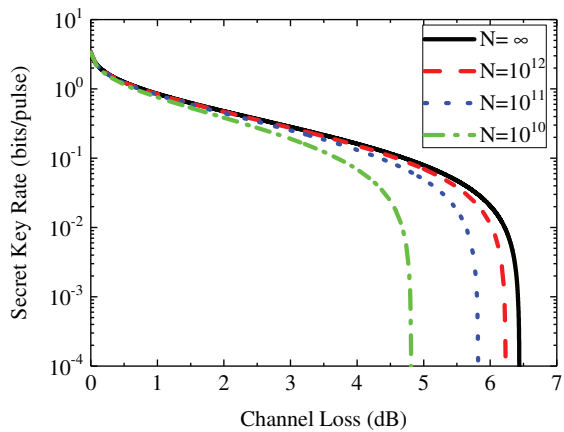}\\
  \caption{(Color online) Secret key rates of squeezed-state CV-MDI QKD protocol against coherent attack in the extremely asymmetric case (${T_{BC}} = 0 $) with reverse reconciliation. The protocol is under ideal modulation variances ${V_A} = {V_B} = {10^5}$ and perfect reconciliation efficiency $\beta  = 1$. The block lengths from left to right curves correspond to $N = {10^{10}}$ (green dot-dashed line), ${10^{11}}$ (blue dot line), ${10^{12}}$ (red dashed line) and $ \infty $ (black solid line), respectively. Here the discretization parameter is set to $d=13$, the excess noise to ${\varepsilon _1} = {\varepsilon _2} = 0.002$, and the overall security parameter is smaller than ${10^{ - 20}}$.}
\label{ideal_asymmetric_R}
\end{figure}

\begin{figure}[t]
  \centering
  \includegraphics[width=7.5cm]{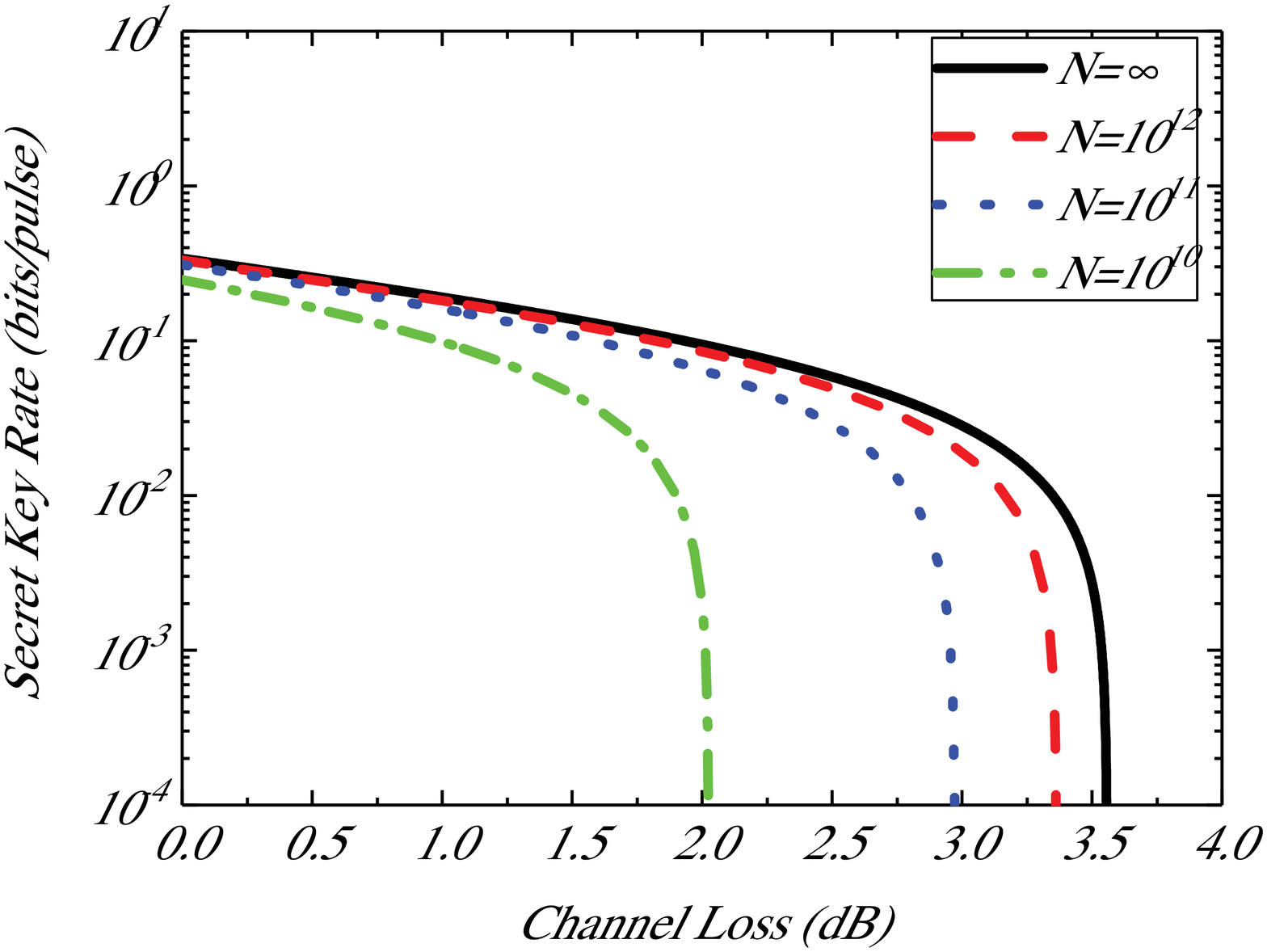}\\
  \caption{(Color online) Secret key rates of squeezed-state CV-MDI QKD protocol against coherent attack in the extremely asymmetric case (${T_{BC}} = 0$) with reverse reconciliation. The protocol is under practical modulation variances ${V_A} = {V_B} = 5.04$ and imperfect reconciliation efficiency $\beta  = 96.9\%$. The block lengths from left to right curves are $N = {10^{10}}$ (green dot-dashed line), ${10^{11}}$ (blue dot line), ${10^{12}}$ (red dashed line) and $ \infty $ (black solid line), respectively. The discretization parameter, excess noises, and security parameters are chosen as in the case of ideal modulation.}
\label{practical_asymmetric_R}
\end{figure}

\begin{figure}[t]
  \centering
  \includegraphics[width=7.5cm]{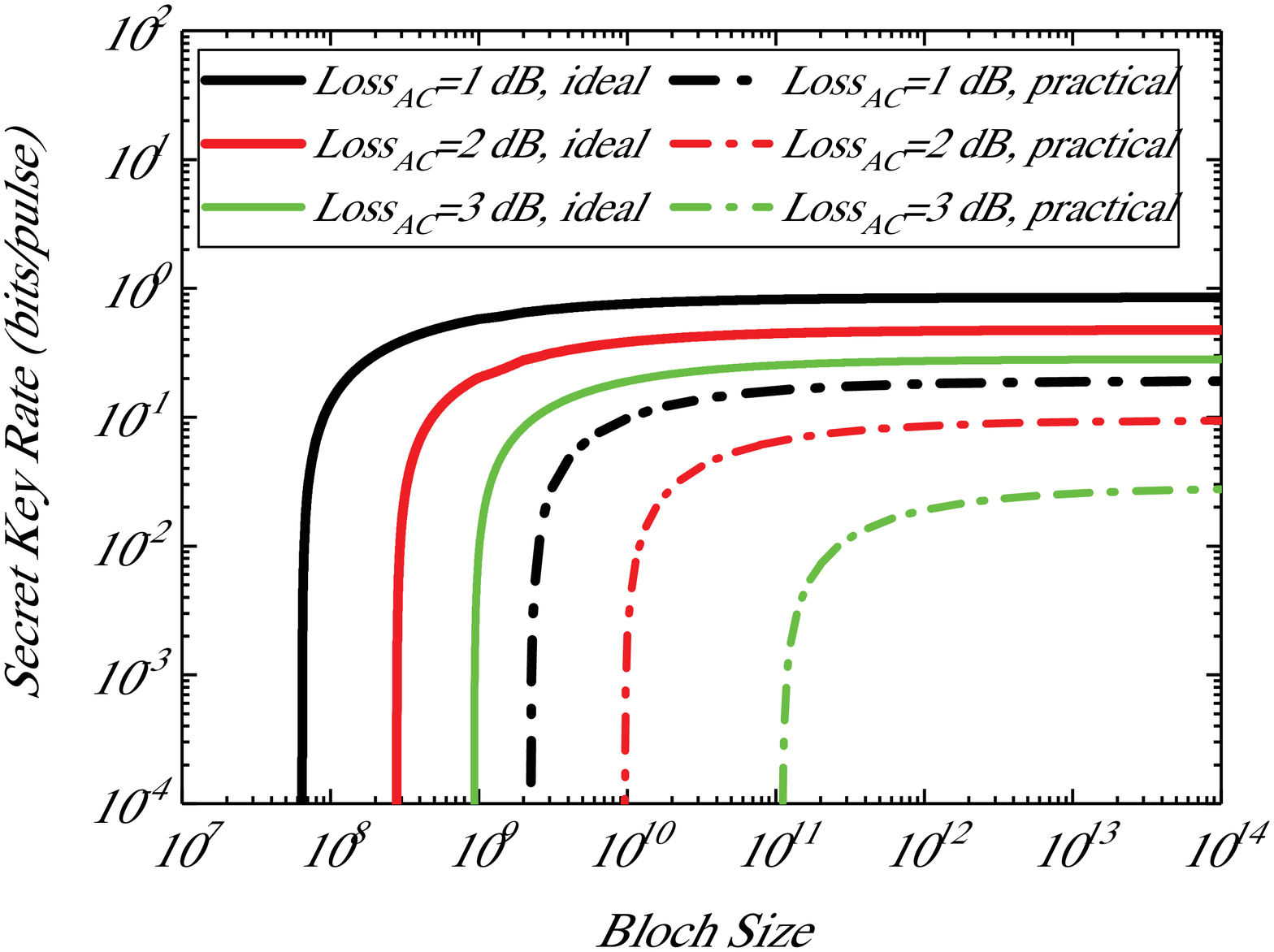}\\
  \caption{(Color online) Secret key rates vs block size for the extremely asymmetric case (${T_{BC}} = 0$) with reverse reconciliation. The solid lines are under the ideal condition that modulation variances ${V_A} = {V_B} = {10^5}$ and perfect reconciliation efficiency $\beta  = 1$. The dot-dashed lines are under the practical condition that practical modulation variances ${V_A} = {V_B} = {5.04}$ and imperfect reconciliation efficiency $\beta  = 96.9\% $. From left to right, the transmittance of the quantum channel corresponds to loss of 1 dB (black line), 2 dB (red line), and 3 dB (green line), respectively.}
\label{N_change_R}
\end{figure}

Similar to the direct reconciliation case, the protocol's performance under reverse reconciliation can be illustrated using the same method. The smooth maximum entropy is the same with that of the direct reconciliation case, while the leakage information is different.

Both ideal cases and practical cases are taken into consideration and the parameters we choose are the same with those of the direct reconciliation cases. Here large variances $ {{V_A} = {V_B} = {{10}^5}} $ are chosen first to illustrate the performance of ideal modulation (Fig. \ref{ideal_asymmetric_R}), then the practical variances ${V_A} = {V_B} = 5.04$ are exploited to show the realistic performance (Fig. \ref{practical_asymmetric_R}). $ {N=10^{10}}$, $ {N=10^{11}}$, $ {N=10^{12}}$, and the asymptotic regime are considered here as well. For a realistic performance of fiber loss 0.2 dB/km, the total loss can be up to 6.5 dB in the ideal condition and 3.6 dB in the practical condition, corresponding to 32.5 km and 18 km respectively. Hence the reverse reconciliation cases could be feasible in metropolitan range communications.

Figure \ref{N_change_R} shows the relation between block size and secret key rate in the extremely asymmetric circumstance. It illustrates that it is in principle possible to generate secret keys for block sizes of ${10^7} - {10^{12}}$ in reverse reconciliation case, depending on channel losses and the required level of security, which is easier to achieve than direct case.

In general, our numerical simulation results show that the protocol can tolerate at most 2.5 dB channel loss with direct reconciliation and 6.5 dB channel loss with reverse reconciliation against coherent attacks in the extremely asymmetric scenario. Meanwhile, the secret key rate is reduced considering practical squeezing parameter and imperfect reconciliation efficiency. Finite-size effect is also discussed apart from asymptotic regime. When the block size is of the order of ${10^7} - {10^{12}}$, one can achieve high secret key rates depending on the channel loss, thus it is practical on the metropolitan scale with current technologies.

\section{Conclusion}
\label{Conclusion}

In this paper, we present a composable security analysis for squeezed-state CV-MDI QKD against general coherent attacks. Its security analysis is derived based on the entanglement-based scheme and a version of state-independent entropic uncertainty relation is exploited to give a lower bound on the conditional smooth min-entropy by trusting Alice's and Bob's devices. Finite size effect is also taken into consideration, and we use two independent entangling cloner attacks to simulate the performance of the method in both direct and reverse reconciliation cases. The simulation results show that, in extremely asymmetric scenarios, the protocol can tolerate 2.5 dB and 0.64 dB channel losses under ideal and practical conditions with direct reconciliation, and 6.5 dB and 3.6 dB channel losses under ideal and practical conditions with reverse reconciliation. An interesting extension to this paper would be to further add the energy test to remove the trusted source assumption.

\begin{acknowledgments}
This work is supported by the National Natural Science Foundation under Grant (Grant No. 61531003) and the National Science Fund for Distinguished Young Scholars of China (Grant No. 61225003), and China Postdoctoral Science Foundation (Grant No. 2018M630116).
\end{acknowledgments}

\begin{appendix}

\section{Estimation of ${t_q}$ and ${t_p}$ in measurement stage}
\label{Estimation of ${t_q}$ and ${t_p}$ in measurement stage}

The usage of ${t_q}$ and ${t_p}$ in Eq. (\ref{rescale}) in the main text is to ensure that the discreted data between Alice and Bob have strong correlation after states passing through channels. In order to guarantee the difference between the data collected by Alice and Bob is small enough, one possible solution is to rescale one of two communicated parties' data such that the second moments of Alice's and Bob's amplitude and phase measurement match.

Supposing Alice and Bob randomly choose amplitude strings $\left\{ {{X_A},{X_B}} \right\}$ of length $m$ and phase strings $\left\{ {{P_A},{P_B}} \right\}$ of length $j$ to estimate parameter ${t_q}$ and ${t_p}$ respectively. Here the estimation of ${t_q}$ is demonstrated as an example, and that of ${t_p}$ can be calculated using the same method.

First, considering the scenario where there is no rescaled and discretization processes in the measurement phase, theoretically the average value of amplitude measurement outcomes both in Alice's and Bob's sides can be estimated by

\begin{equation}\
\hat E\left( {{Q_A}} \right) = \frac{1}{m}\sum\limits_{i = 1}^m {Q_A^i} , \hat E\left( {{Q_B}} \right) = \frac{1}{m}\sum\limits_{i = 1}^m {Q_B^i} ,
\end{equation}
and the variance of amplitude measurement outcomes both in Alice's and Bob's sides can be written as

\begin{equation}\
\hat \sigma \left( {{Q_A}} \right) = \frac{1}{m}\sum\limits_{i = 1}^m {{{\left( {Q_A^i - \hat E\left( {{Q_A}} \right)} \right)}^2}},
\end{equation}
and

\begin{equation}\
\hat \sigma \left( {{Q_B}} \right) = \frac{1}{m}\sum\limits_{i = 1}^m {{{\left( {Q_B^i - \hat E\left( {{Q_B}} \right)} \right)}^2}}.
\label{variance_B}
\end{equation}

After taking rescaled process into account, the estimators of new date ${\tilde Q_A}$ should satisfy the following forms

\begin{equation}\
\hat E\left( {{{\tilde Q}_A}} \right) = \frac{1}{m}\sum\limits_{i = 1}^m {\tilde Q_A^i} ,
\end{equation}

\begin{equation}\
\hat \sigma \left( {{{\tilde Q}_A}} \right) = \frac{1}{m}\sum\limits_{i = 1}^m {{{\left( {\tilde Q_A^i - \hat E\left( {{{\tilde Q}_A}} \right)} \right)}^2}}.
\label{variance_A}
\end{equation}

In order to match the variances of Alice's and Bob's measurement data, the values of Eq. (\ref{variance_B}) and Eq. (\ref{variance_A}) should be the same. When discretization process is done, the parameter ${t_q}$ can be estimated by

\begin{equation}\
{t_q} = \sqrt {\frac{{\sum\limits_{i = 1}^m {{{\left( {X_B^i - \hat E\left( {{X_B}} \right)} \right)}^2}} }}{{\sum\limits_{i = 1}^m {{{\left( {X_A^i - \hat E\left( {{X_A}} \right)} \right)}^2}} }}} ,
\end{equation}
where $\hat E\left(  \cdot  \right)$ is the estimator of the average value of measured data. Therefore, parameter ${t_p}$ can be written using the same estimation method, which reads

\begin{equation}\
{t_p} = \sqrt {\frac{{\sum\limits_{i = 1}^j {{{\left( {P_B^i - \hat E\left( {{P_B}} \right)} \right)}^2}} }}{{\sum\limits_{i = 1}^j {{{\left( {P_A^i - \hat E\left( {{P_A}} \right)} \right)}^2}} }}} .
\end{equation}

 \begin{figure}[b]
  \centering
  \includegraphics[width=7.5cm]{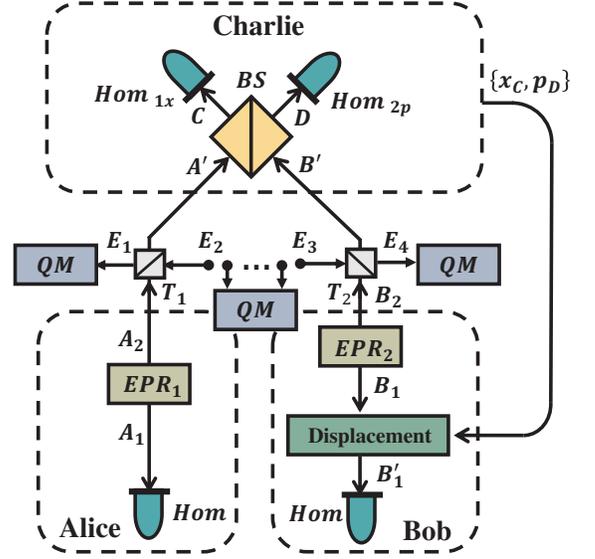}\\
  \caption{(Color online) EB scheme of the squeezed-state CV-MDI QKD protocol with Eve's attacks. After two channels, mode ${A_2}$ becomes $A'$, and mode ${B_2}$ becomes $B'$. QM is the quantum memory.}
\label{entangling_cloner_attacks}
\end{figure}

\section{Derivation of covariance matrixes}
\label{Covariance matrixes and calculating the leakage information in error correction}

Considering the EB version of the squeezed-state CV-MDI QKD protocol, Eve's attacks can be modeled by two independent entangled-cloner attacks (shown in Fig. \ref{entangling_cloner_attacks} and supposing modes ${E_2}$ and ${E_3}$ are independent for simplification), where channel$_{AC}$ and channel$_{BC}$ are replaced by two BS with transmittances ${T _1}$ and ${T _2}$ respectively. The covariance matrixes of two BS can be written as

\begin{equation}\
S_{BS}^{A\left( B \right)} = \left( {\begin{array}{*{20}{c}}
{\sqrt {{T_{1\left( 2 \right)}}} }&{\sqrt {1 - {T_{1\left( 2 \right)}}} }\\
{ - \sqrt {1 - {T_{1\left( 2 \right)}}} }&{\sqrt {{T_{1\left( 2 \right)}}} }
\end{array}} \right),
\end{equation}

After passing two channels, mode ${A_2}$ becomes $A'$, and mode ${B_2}$ becomes $B'$, and the following relationships of quadratures hold

\begin{equation}\
{\hat A' = \sqrt {{T_1}} {{\hat A}_2} + \sqrt {1 - {T_1}} {{\hat E}_2}},
\end{equation}
and
\begin{equation}\
{\hat B' = \sqrt {{T_2}} {{\hat B}_2} + \sqrt {1 - {T_2}} {{\hat E}_3}}.
\end{equation}

Then Charlie applies Bell detection of the measured states. Modes $A'$ and $B'$ are combined with a balanced beam splitter with output modes $C$ and $D$. Therefore we can get modes $C$ and $D$ as

\begin{widetext}
\begin{align}
\hat C  = \frac{1}{{\sqrt 2 }}\left( {\hat A' - \hat B'} \right)
 = \frac{1}{{\sqrt 2 }}\left( {\sqrt {{T_1}} {{\hat A}_2} - \sqrt {{T_2}} {{\hat B}_2}} \right) + \frac{1}{{\sqrt 2 }}\left( {\sqrt {1 - {T_1}} {{\hat E}_2} - \sqrt {1 - {T_2}} {{\hat E}_3}} \right),
\end{align}
and

\begin{align}
\hat D  = \frac{1}{{\sqrt 2 }}\left( {\hat A' + \hat B'} \right)
 = \frac{1}{{\sqrt 2 }}\left( {\sqrt {{T_1}} {{\hat A}_2} + \sqrt {{T_2}} {{\hat B}_2}} \right) + \frac{1}{{\sqrt 2 }}\left( {\sqrt {1 - {T_1}} {{\hat E}_2} + \sqrt {1 - {T_2}} {{\hat E}_3}} \right).
\end{align}

Before Charlie makes a Bell measurement to the $C$ and $D$ modes, the whole state ${\rho _{{A_1}CD{B_1}}}$ can be described by the $8 \times 8$ covariance matrix ${\gamma _{{A_1}CD{B_1}}}$, given by

\begin{equation}\
\resizebox{.9\hsize}{!}{$
{\gamma _{{A_1}CD{B_1}}} = \left( {\begin{array}{*{20}{c}}
{{V_A}{{\mathbb{I}_2}}}&{\sqrt {\frac{1}{2}{T_1}\left( {V_A^2 - 1} \right)} {\sigma _z}}&{\sqrt {\frac{1}{2}{T_1}\left( {V_A^2 - 1} \right)} {\sigma _z}}&{0{{\mathbb{I}_2}}}\\
{\sqrt {\frac{1}{2}{T_1}\left( {V_A^2 - 1} \right)} {\sigma _z}}&{\left( {\frac{1}{2}{T_1}\left( {{V_A} + {\chi _1}} \right) + \frac{1}{2}{T_2}\left( {{V_A} + {\chi _2}} \right)} \right){{\mathbb{I}_2}}}&{\left( {\frac{1}{2}{T_1}\left( {{V_A} + {\chi _1}} \right) - \frac{1}{2}{T_2}\left( {{V_A} + {\chi _2}} \right)} \right){{\mathbb{I}_2}}}&{\sqrt {\frac{1}{2}{T_2}\left( {V_B^2 - 1} \right)} {\sigma _z}}\\
{\sqrt {\frac{1}{2}{T_1}\left( {V_A^2 - 1} \right)} {\sigma _z}}&{\left( {\frac{1}{2}{T_1}\left( {{V_A} + {\chi _1}} \right) - \frac{1}{2}{T_2}\left( {{V_A} + {\chi _2}} \right)} \right){{\mathbb{I}_2}}}&{\left( {\frac{1}{2}{T_1}\left( {{V_A} + {\chi _1}} \right) + \frac{1}{2}{T_2}\left( {{V_A} + {\chi _2}} \right)} \right){{\mathbb{I}_2}}}&{ - \sqrt {\frac{1}{2}{T_2}\left( {V_B^2 - 1} \right)} {\sigma _z}}\\
{0{{\mathbb{I}_2}}}&{\sqrt {\frac{1}{2}{T_2}\left( {V_B^2 - 1} \right)} {\sigma _z}}&{ - \sqrt {\frac{1}{2}{T_2}\left( {V_B^2 - 1} \right)} {\sigma _z}}&{{V_B}{{\mathbb{I}_2}}}
\end{array}} \right)
$},
\end{equation}
where ${\varepsilon _1}$ and ${\varepsilon _2}$ in ${\chi _1} = {1 \mathord{\left/
 {\vphantom {1 {{T_1}}}} \right. \kern-\nulldelimiterspace} {{T_1}}} - 1 + {\varepsilon _1}$, ${\chi _2} = {1 \mathord{\left/{\vphantom {1 {{T_2}}}} \right.\kern-\nulldelimiterspace} {{T_2}}} - 1 + {\varepsilon _2}$ are the excess noises of the corresponding channels.

The measurement results ${x_C}$ and ${p_D}$ are announced by Charlie in a public channel so that Bob can displace mode ${B_1}$ to ${B'_1}$, whose relationships of quadratures read

\begin{align}
{\hat B'_{1x}} = {\hat B_{1x}} + g{\hat C_x}
  = \left( {{{\hat B}_{1x}} - g\sqrt {\frac{{{T_2}}}{2}} {{\hat B}_{2x}}} \right) + g\sqrt {\frac{{{T_1}}}{2}} {\hat A_{2x}} + \frac{g}{{\sqrt 2 }}\left( {\sqrt {1 - {T_1}} {{\hat E}_{2x}} - \sqrt {1 - {T_2}} {{\hat E}_{3x}}} \right),
\end{align}
and

\begin{align}
{\hat B'_{1p}} = {\hat B_{1p}} + g{\hat D_p}
  = \left( {{{\hat B}_{1p}} + g\sqrt {\frac{{{T_2}}}{2}} {{\hat B}_{2p}}} \right) + g\sqrt {\frac{{{T_1}}}{2}} {\hat A_{2p}} + \frac{g}{{\sqrt 2 }}\left( {\sqrt {1 - {T_1}} {{\hat E}_{2p}} + \sqrt {1 - {T_2}} {{\hat E}_{3p}}} \right).
\end{align}

Hence the covariance matrix ${\gamma _{{A_1}{B_1}^\prime }}$ of the state ${\rho _{{A_1}{B_1}^\prime }}$ can be written as Eq. (\ref{gammaAB}) in the main text.

\end{widetext}

\section{Derivation of discrete Shannon entropy}
\label{Derivation of discrete Shannon entropy}
A continuous variable can always be approximated as a discrete variable with finite resolution digital discretization, and the smaller the discreted unit is, the closer the discrete variable is to the continuous variable.

Assuming that the variable $x$ belongs to the interval $x \in \left[ {a,b} \right]$, whose probability density function is denoted by $p\left( x \right)$, we divide this interval into $n$ continuous intervals with same length $\delta $, where $\delta  = \frac{{b - a}}{n}\ $. According to the mean value theorem of integrals, there is a value ${x_i}$ in each interval ${x_i} \in \left[ {a + \left( {i - 1} \right)\delta ,a + i\delta } \right]$, where $i = 1,2,...n$, and ${x_i}$ should satisfy

\begin{equation}\
{p_i} = p\left( {{x_i}} \right)\delta  = \int_{a + \left( {i - 1} \right)\delta }^{a + i\delta } {p\left( x \right)dx} ,
\end{equation}
where ${p_i}$ is the probability in each intervals. Therefore, the discrete Shannon entropy $H\left( {{x^\delta }} \right)$ can be derived by

\begin{align}
H\left( {{x^\delta }} \right) &=  - \sum\limits_{i = 1}^n {{p_i}\log {p_i}} \notag\\
 &=  - \sum\limits_{i = 1}^n {p\left( {{x_i}} \right)\delta \log \left[ {p\left( {{x_i}} \right)\delta } \right]}                   \notag\\
  &=  - \sum\limits_{i = 1}^n {p\left( {{x_i}} \right)\delta \log p\left( {{x_i}} \right) - \left( {\log \delta } \right)\sum\limits_{i = 1}^n {p\left( {{x_i}} \right)\delta } }     \notag\\
   &=  - \sum\limits_{i = 1}^n {p\left( {{x_i}} \right)\delta \log p\left( {{x_i}} \right) - \log \delta }.
\end{align}

Here we use the relation $\sum\limits_{i = 1}^n {p\left( {{x_i}} \right)} \delta  = 1$. The limit of $H\left( {{x^\delta }} \right)$ as $\delta $ approaches zero, goes toward the entropy of continuous variable, which reads

\begin{align}
\mathop {\lim }\limits_{\delta  \to 0} {H_n}\left( {{x^\delta }} \right) &= \mathop {\lim }\limits_{\delta  \to 0} \left[ { - \sum\limits_{i = 1}^n {{p_n}\left( {{x_i}} \right)\delta \log {p_n}\left( {{x_i}} \right) - \log \delta } } \right]       \notag\\
 &=  - \int\limits_a^b {p\left( x \right)\log p\left( x \right)dx}  - \mathop {\lim }\limits_{\delta  \to 0} \left( {\log \delta } \right)   \notag\\
 &\buildrel \Delta \over = {h}\left( x \right) + H\left( \delta  \right),
\end{align}
where ${h}\left( x \right) =  - \int\limits_a^b {p\left( x \right)\log p\left( x \right)dx}$ denotes the differential entropy, and $H\left( \delta  \right) = -\mathop {\lim }\limits_{\delta  \to 0} \left( {\log \delta } \right)$.

Now consider a normal distribution

\begin{equation}\
g\left( x \right) = \frac{1}{{\sqrt {2\pi } \sigma }}\exp \left( { - \frac{{{x^2}}}{{2{\sigma ^2}}}} \right),
\end{equation}
with variance ${\sigma ^2}$. The differential entropy of a normal distribution reads,

\begin{align}
h\left( x \right) &=  - \int {g\left( x \right)} \log \left[ {g\left( x \right)} \right]   \notag\\
 &=  - \int {dx} g\left( x \right)\left( { - \frac{{{x^2}}}{{2{\sigma ^2}}} + \frac{1}{2}\log \left( {2\pi {\sigma ^2}} \right)} \right)   \notag\\
 &= \frac{1}{2} + \frac{1}{2}\log \left( {2\pi {\sigma ^2}} \right)     \notag\\
 &= \log \left( {\sqrt {2\pi e {\sigma ^2}} } \right).
\end{align}

Supposing another continuous variable $y \in \left[ {a,b} \right]$, the relationships $\int {p\left( y \right)dy = 1}$ and $\int {p\left( {{x \mathord{\left/{\vphantom {x y}} \right.\kern-\nulldelimiterspace} y}} \right)dx = 1} $ hold, where $p\left( {{x \mathord{\left/{\vphantom {x y}} \right.\kern-\nulldelimiterspace} y}} \right)$ is the conditional probability density function of $x$ given $y$. Then

\begin{align}
H\left( {{{{x^\delta }} \mathord{\left/{\vphantom {{{x^\delta }} {{y^\delta }}}} \right. \kern-\nulldelimiterspace} {{y^\delta }}}} \right) =  - \sum\limits_j {p\left( {{y_j}} \right)\delta } \sum\limits_i {p\left( {{{{x_i}} \mathord{\left/{\vphantom {{{x_i}} {{y_j}}}} \right.
 \kern-\nulldelimiterspace} {{y_j}}}} \right)\delta } \log \left[ {p\left( {{{{x_i}} \mathord{\left/{\vphantom {{{x_i}} {{y_j}}}} \right.\kern-\nulldelimiterspace} {{y_j}}}} \right)\delta } \right]    \notag\\
 =  - \sum\limits_j {p\left( {{y_j}} \right)\delta } \sum\limits_i {p\left( {{{{x_i}} \mathord{\left/
 {\vphantom {{{x_i}} {{y_j}}}} \right.
 \kern-\nulldelimiterspace} {{y_j}}}} \right)\delta } \log \left[ {p\left( {{{{x_i}} \mathord{\left/
 {\vphantom {{{x_i}} {{y_j}}}} \right.
 \kern-\nulldelimiterspace} {{y_j}}}} \right)} \right] - \log \delta ,
\end{align}
and the limit of $H\left( {{{{x^\delta }} \mathord{\left/{\vphantom {{{x^\delta }} {{y^\delta }}}} \right.
 \kern-\nulldelimiterspace} {{y^\delta }}}} \right)$ as $\delta $ approaches zero reads

\begin{equation}\
\mathop {\lim }\limits_{\delta  \to 0} H\left( {{{{x^\delta }} \mathord{\left/
 {\vphantom {{{x^\delta }} {{y^\delta }}}} \right.
 \kern-\nulldelimiterspace} {{y^\delta }}}} \right) = h\left( {{x \mathord{\left/
 {\vphantom {x y}} \right.
 \kern-\nulldelimiterspace} y}} \right) + H\left( \delta  \right),
\end{equation}
where $h\left( {{x \mathord{\left/ {\vphantom {x y}} \right.\kern-\nulldelimiterspace} y}} \right)$ is the conditional entropy of $x$ given $y$, which reads

\begin{equation}\
h\left( {{x \mathord{\left/
 {\vphantom {x y}} \right.
 \kern-\nulldelimiterspace} y}} \right) =  - \int {\int {p\left( y \right)p\left( {{x \mathord{\left/
 {\vphantom {x y}} \right.
 \kern-\nulldelimiterspace} y}} \right)\log p\left( {{x \mathord{\left/
 {\vphantom {x y}} \right.
 \kern-\nulldelimiterspace} y}} \right)} } dxdy.
\end{equation}

Hence the mutual information $I\left( {{x^\delta }:{y^\delta }} \right)$ between discrete variables ${{x^\delta }}$ and ${{y^\delta }}$, as approaches zero, can be approximated,

\begin{align}
\mathop {\lim }\limits_{\delta  \to 0} I\left( {{x^\delta }:{y^\delta }} \right) &= \mathop {\lim }\limits_{\delta  \to 0} \left[ {H\left( {{x^\delta }} \right) - H\left( {{{{x^\delta }} \mathord{\left/
 {\vphantom {{{x^\delta }} {{y^\delta }}}} \right.\kern-\nulldelimiterspace} {{y^\delta }}}} \right)} \right] \notag\\
 &= h\left( x \right) - \log \delta  - \left( {h\left( {{x \mathord{\left/
 {\vphantom {x y}} \right.\kern-\nulldelimiterspace} y}} \right) - \log \delta } \right)    \notag\\
 &= I\left( {x:y} \right),
\end{align}
where $I\left( {x:y} \right)$ is the mutual information between continuous variables $x$ and $y$. Therefore, if the length of intervals $\delta $ is small enough, we can always regard the continuous variable mutual information $I\left( {x:y} \right)$ as the approximation of the discrete one.

\section{Symmetric cases under two correlated modes attacks and the comparison with the PLOB bound  }
\label{The protocol under different transmission losses}

In the main text, the performance of squeezed-state CV-MDI QKD under coherent attacks has been discussed focusing on the extremely asymmetric cases ($T_{BC}=0$). However, in the calculation of security key rate, all of effects caused by the channel are usually treated as Eve's contribution. The optimal attack strategy provide the maximum information for Eve to reduce the security of the protocol to the greatest extent, so it is important to study the protocol under more general attack strategies.

\begin{figure}[t]
  \centering
  \includegraphics[width=7.5cm]{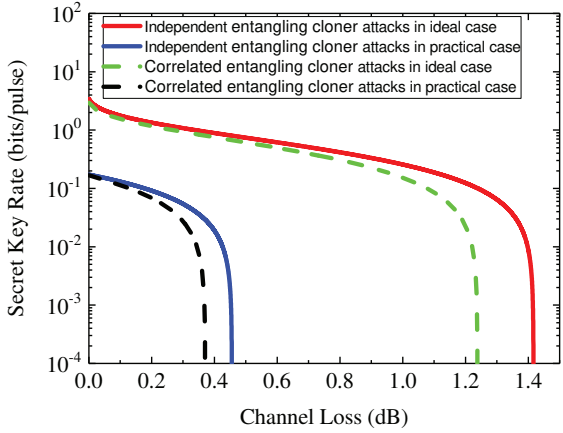}\\
  \caption{(Color online) Comparison of the secret key between independent entangling cloner attacks model and correlated modes attacks model in symmetric case. The ideal case is under ideal modulation variances ${V_A} = {V_B} = {10^5}$ and perfect reconciliation efficiency $\beta  = 1$. The practical case is with practical modulation variances ${V_A} = {V_B} = 5.04$ and imperfect reconciliation efficiency $\beta  = 96.9\% $. The solid lines are the secret key rates under two independent entangling cloner attacks model. The dashed lines are the secret key rates under two correlated modes attacks model.}
\label{two_correlated_mode}
\end{figure}

According to Ref. \cite{Phys.Rev.A.91.022320.2015}, the covariance matrix of two correlated modes ${E_2}$ and ${E_3}$ measured by Eve (Fig. \ref{entangling_cloner_attacks}) has the following form:

\begin{equation}\
{\gamma _{{E_2}{E_3}}} = \left( {\begin{array}{*{20}{c}}
{{V_{{E_2}}}{\mathbb {I}}_2}&G\\
G&{{V_{{E_3}}}{\mathbb {I}}_2}
\end{array}} \right),
\end{equation}
where $G$ is the correlation term. Supposing ${V_{{E_2}}} = {V_{{E_3}}} = {V_E}$, to achieve maximum correlation between Eve's modes, $G$ is chosen as $\sqrt {V_E^2 - 1} {\sigma _z}$ due to the uncertainty principle, and the final covariance matrix of ${\rho _{{A_1}{B_1}^\prime }}$ under two correlated mode attack model is given by

\begin{equation}\
{\gamma _{{A_1}{B_1}^\prime }} = \left( {\begin{array}{*{20}{c}}
{{V_A}{{\mathbb {I}}_2}}&{\sqrt {T\left( {V_A^2 - 1} \right)} {\sigma _z}}\\
{\sqrt {T\left( {V_A^2 - 1} \right)} {\sigma _z}}&{\left[ {T\left( {{V_A} - 1} \right) + 1 + Te'} \right]{{\mathbb {I}}_2}}
\end{array}} \right).
\end{equation}

The equivalent excess noise $e'$ reads

\begin{align}\
e' &= 1 + \frac{1}{{{T_1}}}\left[ {2 + {T_2}\left( {{\varepsilon _2} - 2} \right) + {T_1}\left( {{\varepsilon _1} - 1} \right)} - {C_E}\right]          \notag\\
&+ \frac{1}{{{T_1}}}{\left( {\frac{{\sqrt 2 }}{g}\sqrt {{V_B}-1}  - \sqrt {{T_2}} \sqrt {{V_B} + 1} } \right)^2},
\end{align}
where ${C_E} = \frac{2}{{{T_1}}}\sqrt {\left( {1 - {T_1}} \right)\left( {1 - {T_2}} \right)} \left\langle {{E_{2x}}{E_{3x}}} \right\rangle$ is the noise contribution of $x$-quadrature induced by the correlation of Eve's two modes, and ${C_E} = -\frac{2}{{{T_1}}}\sqrt {\left( {1 - {T_1}} \right)\left( {1 - {T_2}} \right)} \left\langle {{E_{2p}}{E_{3p}}} \right\rangle$ is the corresponding noise contribution of $p$-quadrature.

If one of $T_1$ and $T_2$ is equal to zero, corresponding to the extremely asymmetric case (discussed in our main text), the contribution of the two modes' correlation in Eve's attack disappears, and the independent entangling cloner attack and two correlated mode attack are equivalent in this situation. Therefore, the model of two independent entangling cloner attack can simplify our numerical simulation in the main text.

Under correlated mode attacks, the secret key rate formula is the same as that in the main text (Eq. \ref{keyrate}), but compared with independent entangling cloner attacks, the cases that both $T_{1}$ and $T_{2}$ are not equal to zero will increase the leakage information and then decrease the key rate in the two correlated mode attack model. Typically, in the symmetric case, where the relay is located in the middle of Alice and Bob, we compare the results between two independent entangling cloner attacks and the two correlated modes attacks in Fig. \ref{two_correlated_mode}, and the discussion is under the asymptotic regime.

\begin{figure}[t]
  \centering
  \includegraphics[width=7.5cm]{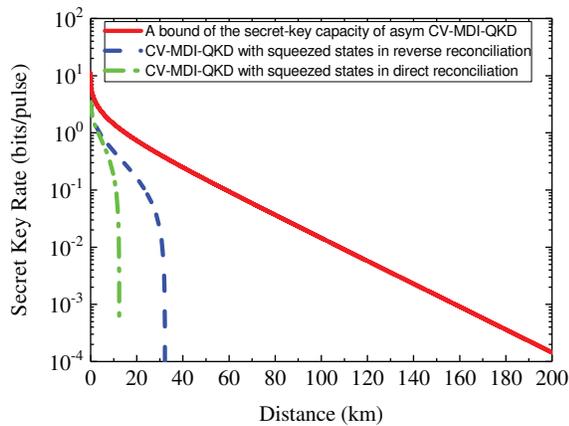}\\
  \caption{(Color online) Comparison of the secret key between the extremely asymmetric results and the PLOB bound. Our protocol is under ideal modulation variances ${V_A} = {V_B} = {10^5}$ and perfect reconciliation efficiency $\beta  = 1$. The red solid line is the bound of secret key capacity of asymmetric CV-MDI-QKD. The green dot-dashed line and the blue dashed line are the secret key rates of CV-MDI-QKD with squeezed states in direct and reverse reconciliation cases respectively.}
\label{comparation}
\end{figure}

It can be seen that in the symmetric case, which is the worst case for the two correlated modes attacks (because the correlation term ${C_E}$ reaches the maximum value), this correlated attack model will slightly degrade the performance of the protocol. Moreover, the more asymmetric the protocol, the smaller the impact of the correlated attack on the secret key rate.

In experiment, the covariance matrices can be obtained by data statistics, so there is no need to assume which model Eve's attack strategy belongs to before the protocol starts. If Eve's attack is stronger than the model we give in our simulation, the correlation between Alice and Bob's data will decline, so the estimation of max-entropy will increase, causing the decrease of the min-entropy. Moreover, Alice and Bob need to sacrifice more keys to do the error correction in classical post-processing process, and it will leak more information to Eve. Therefore, if the two correlated mode attack is exploited by Eve, the secret key rate can still be calculated using Eq. \ref{keyrate} and it will not influence the security analysis of the protocol, but the secret key rate will decrease.

We also compare the extremely asymmetric results with the PLOB bound \cite{Nat.Commun.8.15043.2017} shown in Fig. \ref{comparation}, which is the secret-key capacity of the lossy channel. It can be seen that even though there is still a gap between the secret key rate of our protocol and the key capacity bound above, the final key rate can be improved using current technologies, such as photon subtraction method \cite{Phys.Rev.A.93.012310.2016, Quantum.Inf.Process.16.184.2017, arXiv.1711.04225.2017} and adding trusted noise method \cite{PhysRevA.90.052325, Phys.Rev.Lett.102.130501.2009}.

\end{appendix}
~

%\bibliographystyle{unsrt}
%\bibliography{reference}

\begin{thebibliography}{99}
\bibitem{note} Note that in general, the smooth min-entropy $H_{\min }^\varepsilon \left( {{X_A}|E} \right)$ depends on the selection of information reconciliation methods. However, in our analysis the estimation of max-entropy is reduced by the function $n\log \gamma \left( {{d_0} + \mu } \right)$, which is the same as reverse reconciliation case, and only depends on the collected data between Alice and Bob. Therefore, we assume that $H_{\min }^\varepsilon \left( {{X_A}|E} \right)$ is unrelated to the reconciliation methods in our paper.
\end{thebibliography}

\bibliographystyle{unsrt}

\end{document}